\documentclass[pra,twocolumn]{revtex4-1}
\usepackage[colorlinks,bookmarks=false,citecolor=blue,linkcolor=red,urlcolor=black]{hyperref}
\usepackage{graphicx}
\graphicspath{ {./graph/} }
\usepackage{amssymb,amsmath} 
\usepackage{amsfonts}
\usepackage{color}
\usepackage[utf8]{inputenc}
\usepackage[inline]{enumitem}
\usepackage[T1]{fontenc}
\usepackage{epsfig}
 
 \def\urlprefix{}
 \def\url#1{}
\def\be{\begin{equation}}
\def\ee{\end{equation}}
\def\bea{\begin{eqnarray}}
\def\eea{\end{eqnarray}}
\def\bi{\begin{itemize}}
\def\ei{\end{itemize}}
\def\bin{\begin{enumerate}}
\def\ein{\end{enumerate}}

\begin{document}
\title{The extended states in disordered 1D systems in the presence of the generalized $N$-mer correlations}
\author{Jan Major} 
\affiliation{\mbox{Instytut Fizyki imienia Mariana Smoluchowskiego, Uniwersytet Jagiello\'nski, \L{}ojasiewicza 11, 30-348 Krak\'ow, Poland}}

\begin{abstract}
We have been investigating the problem of the Anderson localization in a disordered one dimensional tight-binding model. The disorder is created by the interaction of mobile particles with other species, immobilized at random positions.  We introduce a novel method of creating correlations in the optical lattices with such a kind of disorder by using two different lattices with commensurate lattice lengths to hold two species of the particles. Such a model, called the generalized random $N$-mer model leads to the appearance of multiple extended states in contrary to a localization of all states usually expected in one dimension. We develop a method, based on properties of transfer matrices which can be used to determine the presence of extended states and their energies for that class of correlations. Analytical results are compared with the numerical calculations for several cases which can be realized in cold-atom experiments.
\end{abstract}
\maketitle
\section{Introduction}
The phenomenon of the Anderson Localization (AL) -- a suppression of transport in disordered systems -- has been extensively studied theoretically for over fifty years \cite{anderson58,Lagendijk2009}. In a disordered medium a wavefunction of a particle, after multiple random scatterings, interferes destructively and as a result the particle localizes with exponential profile. In three dimensional systems the localization appears only above some critical amplitude of the disorder, thus the metal-insulator transition is present, on the other hand in one dimension, the localization occurs always -- even for an arbitrarily weak disorder \cite{abrahams1979}. An important exception is the case of correlated disorder where extended states can appear even in 1D. As those states are separated and do not form a continuous conducting phase, there is no phase transition. Nevertheless, if the system is finite, transport is possible in the windows of energy around extended states \cite{DRDM,Moura1998,Piraud2013,Caetano2005,Zhao2007,vignolo2010}.

As AL is a one-particle interference effect, the strong electron-electron and electron-phonon interactions make it impossible to observe it in the solid-state systems, for which it has been originally devised. A medium, which allows investigation of AL as well as many other solid-state physics models, are ultracold atomic gases in electromagnetic fields. Particularly, periodic structures as crystals can be simulated with optical lattices allowing unprecedented tunability of parameters, especially when supported with techniques as the fast periodic modulation enabling a modification of tunneling amplitudes or the Feshbach resonance allowing tuning of interaction strengths \cite{Bloch2008,Jaksch2005,Lewenstein12,holthaus2005,Lignier2007,Struck2011,Chin2010,targonska2012,Przysiezna2015}. Thus the optical lattices are thought as the best up-to-date realization of the idea of quantum simulator -- a system easy to control experimentally which is mimicking physics of other system hard to investigate.

AL has been observed in the ultracold atomic systems in several setups (for the first time  8 years ago) \cite{billy08,roati2008,anderson3D}. One-particle limit, crucial for the localization has been obtained, whether by using small density of particles or by turning interactions down with the Feshbach resonances. One way of creating disordered potential is the usage of the second species of atoms, immobilized in the lattice at random positions and acting as a static potential, which in the simplest setup realizes as diagonal disorder with values taken from a binary uncorrelated distribution \cite{Castin_2005_binary_disorder,Massignan06,Bongs_2006,horstmann2007,Graham2008}. Although in an experiment correlations could appear as a result of specific preparation procedure, it is hard to create more complex forms of them.

The main aim of this work is to find a class of disordered systems which, being experimentally realizable, exhibits nonstandard localization properties i.e. for most energies atoms are strongly localized but for several resonant energies they travel freely throughout the system.
Such setups could be treated as tuneable band-pass filters for particle energies, trapping all but few atoms with precisely specified energies. Thus, they could be used to shape matter waves leaving a system in a way similar to shaping light pulses described in \cite{Gerber2008}.

Starting point for a study was random $N$-mer model, which is well described theoretically and provides many delocalized modes. However, in the experiment only short chains, giving one or two extended states, could be easily generated. We investigated the extension of the $N$-mer model into the generalized $N$-mer model, which seems to be easier to create in the experiment and also exhibits many delocalized modes. We also provide a description of the transport properties of a wide class of systems (including the N-mer model and its generalization), giving better insight into reasons of appearance of extended states and allowing analytical determination of their energies.

The article is structured as follows: first, we describe the model we are using: one dimensional tight-binding with the disordered on-site energies and optionally with renormalized tunneling amplitudes obtained via a fast periodic modulation of on-site energies. We introduce also correlations in the form of generalized $N$-mers. Next, we depict the general method used for a determination of extended states energies. Finally, we present the method of creating generalized $N$-mers in optical lattices and apply our method to several experimentally realizable systems. We show the numerical results confirming the existence of resonances and discuss viability of used approximations for case of long range interactions. 
\section{Model \label{sec:model}}
Let us consider a one dimensional tight-binding model of non-interacting bosons in an optical lattice described by Hamiltonian (with $\hbar$ set to $1$):
\begin{align}\label{eq:H0}
  H= \sum_i\left(\epsilon_i n_i-t_i( a^\dagger _i a_{i+1}+\mathrm{h.c.})\right),
\end{align}
where $n_i$ is the particle number operator at site $i$, $a_i(a_i^\dagger)$ denotes bosonic annihilation (creation) operators, while $\epsilon_i$ is the on-site energy and $t_i$ the tunneling amplitude between sites $i$ and $i+1$. The disorder in on-site energies $\epsilon_i$ can be introduced by using two species of atoms repulsively interacting with each other (with intraspecies interaction still set to zero). The first fraction, called \emph{mobile}, has non-vanishing tunnelings, whereas the second kind, called \emph{frozen} (denoted with $f$ superscript), is immobilized in deep lattice and acts as the source of static, randomly distributed potential. Such a system can be realized for example by using two overlapping lattices with different polarization and atoms in two hyperfine states each \emph{seeing} only one of the lattices \cite{Castin_2005_binary_disorder} (lattice constants of those two lattices will be denoted $b$ and $b^f$, lattice for \emph{mobile} particles is called main lattice and all sites numbering is made referring to it).
This general scheme can lead to a wide range of different models as one can choose the density of \emph{frozen} particles, $b/b^f$ ratio and the range of interspecies interactions.

An inhomogeneity in tunnelings $t_i$ is present from the bare fact that on-site energies are different (precise calculations can be found in \cite{vignolo2010}), but this effect in our case is negligible. A significant off-diagonal disorder can be introduced into system, by a fast periodic modulation of on-site energies $\epsilon_i\rightarrow\epsilon_i(t)=\epsilon_i + \delta\epsilon_i \sin(\omega t)$, where $\epsilon_i$ is the mean energy and $\delta\epsilon_i$ the amplitude of modulation (in our model it can be obtained by modulation of interspecies interaction). In the scope of the Floquet theory, if only modulation frequency is much higher than the other energy scales in the system (as the tunneling amplitude), one can find a time independent effective Hamiltonian, which governs dynamics of the system for times much larger than modulation period \cite{floquet1883,Shirley1965,holthaus2005,bukov15}. In the presented case, the result is well known \cite{kosior15}, one gets Hamiltonian with renormalized tunnelings:
\begin{align}\label{eq:tunneling}
t_i = \mathcal{J}_0\left(\frac{\delta\epsilon_{i+1}-\delta\epsilon_i}{\omega}\right),
\end{align}
where $\mathcal{J}_0$ is the zeroth order Bessel function (we used unperturbed tunneling rate as the energy scale).

\begin{figure}
\begin{center}
 \includegraphics[width=0.47\textwidth]{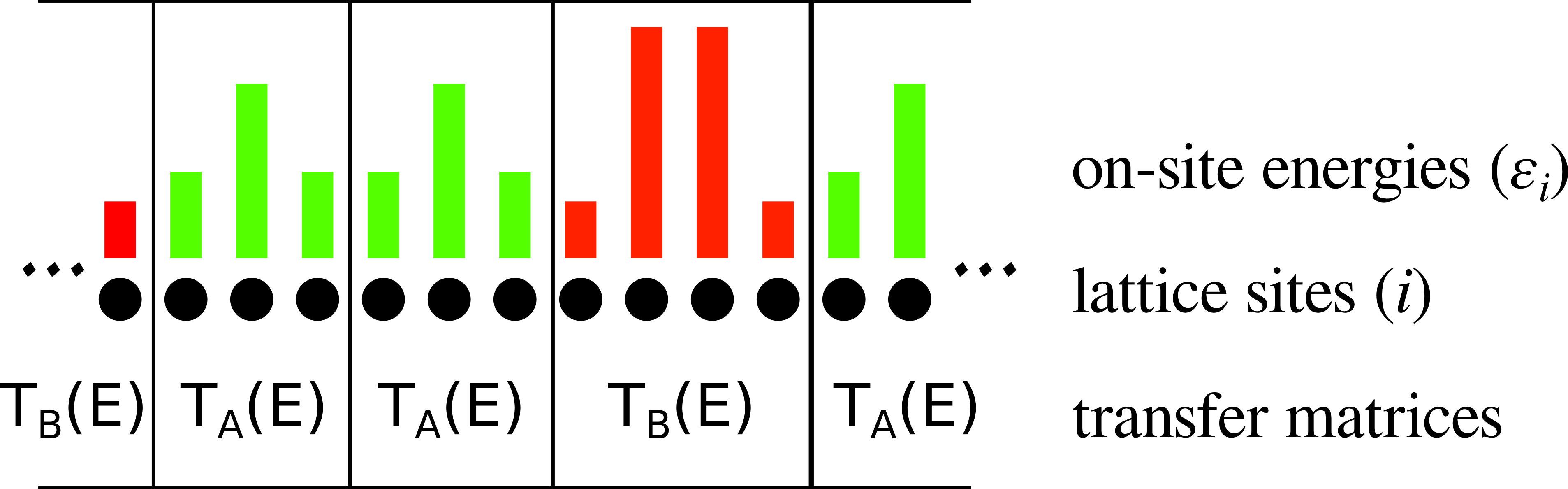}
 \caption{One dimensional lattice with disorder in form of generalized N-mers (gnmers). Boxes denotes different blocks, for each block type set of on-site energies and tunnelings is always the same, so it is possible to calculate transfer matrix through block $T(E)$, for particle with energy $E$. The whole system is composed of randomly ordered blocks\label{pic:GNMER}.}
\end{center}
\end{figure}

Due to the Anderson theory of localization, if $\epsilon_i$ or $t_i$ are taken from a random uncorrelated distribution, all states of the Hamiltonian should be exponentially localized (with exception of a somewhat special case of the band center for a purely off-diagonal disorder \cite{Soukoulis1981}). As it was shown in numerous works, extended states could appear if the disorder distribution is correlated \cite{Piraud2013,DRDM,Moura1998,Caetano2005,Zhao2007,vignolo2010}. In this paper one special class of correlations will be considered: generalized N-mers (gnmers). In this type of correlations we assume that the system is composed of a finite number of different block types (ranging over many lattice sites). Internal structure of each block type is fixed, only their ordering is random (see fig. \ref{pic:GNMER}).

\section{Method of finding extended states\label{sec:method}}
In the system given by the Hamiltonian \eqref{eq:H0}, transport through the site $i$ can be described by a transfer matrix $T_i$ defined by equation:
\begin{align}\label{eq:TM1}
\left(\begin{array}{c}\psi_{i+1}\\ \psi_{i}\end{array}\right)=\left(\begin{array}{c c}\frac{\epsilon_i-E}{t_i} & -\frac{t_{i-1}}{t_i}\\ 1&0\end{array}\right)\left(\begin{array}{c}\psi_{i}\\ \psi_{i-1}\end{array}\right)
\equiv T_i\left(\begin{array}{c}\psi_{i}\\ \psi_{i-1}\end{array}\right),
\end{align}atom monochromatic gun
where $E$ is an energy of a state and $\psi_i$ is a value of a wavefunction on site $i$. By iterating the procedure \eqref{eq:TM1} one can get the transfer matrix describing transport through an arbitrarily chosen part of the system (from site $i$ to $j$)
\begin{align}
 \left(\begin{array}{c}\psi_{j+1}\\ \psi_{j}\end{array}\right)= T_j\cdot\ldots\cdot T_i \left(\begin{array}{c}\psi_{i}\\ \psi_{i-1}\end{array}\right)
\equiv T_i^j \left(\begin{array}{c}\psi_{i}\\ \psi_{i-1}\end{array}\right).
\end{align}
Alternatively, one can focus not on a specific position in the lattice, but rather on one isolated block -- a specific set of on-site energies $\mathbf{e}_X=\{\epsilon^X_1, \ldots,\epsilon^X_{l_X}\}$ and tunnelings $\mathbf{t}_X=\{1,t^X_1, \ldots,t^X_{l_X-1},1\}$ where $l_X$ is length of block. We assume that tunnelings at block edges are equal to one and same for all block types -- we will show later that this assumption is reasonable for considered systems. The transfer matrix for given structure reads:
\begin{align}
T_A(E)=T(\mathbf{e}_A,\mathbf{t}_A,E)=\prod_i^{l_A}\left(\begin{array}{c c}\frac{\epsilon^A_i-E}{t^A_i} & -\frac{t^A_{i-1}}{t^A_i}\\ 1&0\end{array}\right).
\end{align}
For case of the gnmers the transfer matrix for the whole system $\mathcal{T}$ can be decomposed into product of the transfer matrices for different blocks (as on fig. \ref{pic:GNMER}):
\begin{align}
&\mathcal{T}(E)=\prod_X T_X(E),\nonumber\\ &T_X(E)\in\mathbb{T}=\{T_A(E),T_B(E)\ldots\} - \mathrm{finite\; set}.
\end{align}
If it is possible to find energy $E_R$, for which all transfer matrices commute pairwise:
\begin{align}
[T_X(E_R),T_Y(E_R)]=0 \quad \forall T_X(E_R),T_Y(E_R) \in \mathbb{T}, 
\end{align}
the state with energy $E_R$ will be extended. For this energy one matrix diagonalizing all transfer matrices in $\mathbb{T}$ could be found (denoted $C$):
\begin{align}
 \mathcal{T}&= C\left(C^{-1}\mathcal{T}C\right)C^{-1}\nonumber\\&=C\left(\prod_i C^{-1} T_i(E_R) C\right)C^{-1}= C\left(\prod_i D_i(E_R)\right) C^{-1},
\end{align}
where $D_i(E)$ is diagonal form of $T_i(E)$. As long as $E_R$ lies within band for all types of blocks,
\begin{align}
D_i(E_R)=\left(\begin{array}{c c} \alpha_i & 0\\ 0&\alpha_i^*\end{array}\right),
\end{align}
and $|\alpha_i|=1$  $\forall i$. Subsequent multiplication of $D_i$-s preserves the length of $\alpha$ changing its phase. It is a behavior characteristic for a plane wave propagation rather than for the Anderson localization, when we will expect the exponential growth and fall of eigenvalues \cite{delande2011}.

In the further analysis we will restrict ourselves to the case of two structures $T_A(E)$ and $T_B(E)$. Then there is only one commutator and we search for $E_R$ for which:
\begin{align}\label{eq:comm}
[T_A(E_R),T_B(E_R)]=0. 
\end{align}
In this case, even if some parameters change slightly, if only a perturbation is weak enough, $E_R$ could change but the resonance will not vanish. On the other hand, although, it is in principle possible to construct an arbitrary large set of $2\times2$ matrices commuting pairwise, it is much harder to observe resonance then. In an ideal case resonant energy will be the same for every commutator $E_r^{(j)}=E_R \quad\forall j$, but in the perturbed system those energies could change differently: $E_r^{(j)\prime}=E_r^{j}+\varepsilon_j$ and the resonance will vanish. This can be possibly method of making some very precise measurements as we get strong signal only for fine tuned parameters, but in this paper we will not continue this thread.

Equation \eqref{eq:comm} could be simplified if the distribution of on-site energies and tunneling amplitudes in both structures is symmetric (which is the case discussed here but does not have to be true for example for ratchet potentials \cite{Schiavoni2003}), then:
\begin{align}\label{eq:res2rel}
\frac{T_A^{(11)}-T_A^{(22)}}{T_B^{(11)}-T_B^{(22)}}=\frac{T_A^{(21)}}{T_B^{(21)}}
\end{align}
where $T_X^{(ij)}$ denotes the element $ij$ of the matrix $T_X$.

\section{Possible realizations\label{sec:realizations}}
As we have stated, there are numerous scenarios that can be simulated by the general scheme sketched above, here we will focus only on a few of them. Let us assume that at most one \emph{frozen} particle is allowed on each site, which can be realized by using spin polarized fermions or strongly repelling (hardcore) bosons. From now on we fix the energy scale by setting unperturbed tunneling rate to $1$ ($t_i=1$, for all $i$ unless changed, for example by fast periodic modulation as in \eqref{eq:tunneling}).
\paragraph{Binary disorder (bd)} can be created if the lattice constants are equal ($b=b^f$) and the species interact only on-site. Then the $\epsilon_i$ in \eqref{eq:H0}:
\begin{align}
  \epsilon_i^\mathrm{bd}=V n^f_i,forming
\end{align}
where $V$ is the interspecies interaction energy, thus effective on-site energies could take two randomly distributed values $0$ and $V$. In principle, the distribution of \emph{frozen} particles is uncorrelated and no extended modes appear, yet there exists a simple and theoretically well described form of correlations that can exist in such systems -- $N$-mers. In a random $N$-mer model we assume that the \emph{frozen} particles always come in  rows of a given length. Although, it is possible to create this kind of correlations by careful setup of an experiment, as described in \cite{Sedrakyan2011} for the case of random dimers or in \cite{vignolo2010} for the similar case of dual random dimers (frozen particles are always separated by at least one empty site), generating longer chains using such methods is cumbersome.
The analytical expression for the positions of delocalized modes in $N$-mer is well known and can be found using several methods \cite{Peng2004,Zhao2007,Barros2011,Evangelou1993,kosior15b}. It is, however, worth checking how our method works with it and what additional information we can obtain.
By $T^l_\epsilon(E)$ we denote the transfer matrix through the block of length $l$, with on-site energy $\epsilon$. It can be shown that such a matrix reads:
\begin{align}
 T^l_\epsilon(E)=\left(\begin{array}{cc}U_l(\varepsilon/2)&-U_{l-1}(\varepsilon/2)\\U_{l-1}(\varepsilon/2)&-U_{l-2}(\varepsilon/2) \end{array}\right),
\end{align}
where $U_l$ is the $l$-th Chebyshev polynomial of the second kind and $\varepsilon=\epsilon-E$ \cite{kosior15b}. If we have a system composed of rows with on-site energy $\epsilon$ ($T_A=T^l_\epsilon(E)$) and empty spaces of arbitrary length ($T_B=T^1_0(E)$), it is straightforward to check that $[T^l_\epsilon,T^1_0]=0$ if $U_{l-1}=0$. Knowing zeros of Chebyshev polynomial we can get resonant energies:
\begin{align}\label{eq:resnmer}
 E_R=\epsilon+2\cos\left(\frac{\pi}{l}i\right),\quad\mathrm{for}\; i\in\{1,\ldots,l-1\}.
\end{align}
For us the fact that for those energies $T^l_\epsilon(E_R) =\mathbb{I}$ is more important. The identity matrix commutes with everything, therefore in systems, in which one of structures is $N$-meformingr, we will always obtain a set of resonant energies given by \eqref{eq:resnmer} regardless of the other structure. In such a case we can also use recurrence properties of the Chebyshev polynomials ($U_{l+1}(x) = 2xU_l(x)-U_{l-1}(x) $) to simplify the relation \eqref{eq:res2rel} into: 
\begin{align}\label{eq:rescond}
T_B^{(11)}(E)=\epsilon T_B^{(21)}(E)+T_B^{(22)}(E),
\end{align}
where $T_B$ is a transfer matrix for other (not necessarily $N$-mer) type of structure.
\begin{figure}
 \begin{center}
  \includegraphics[width=0.45\textwidth]{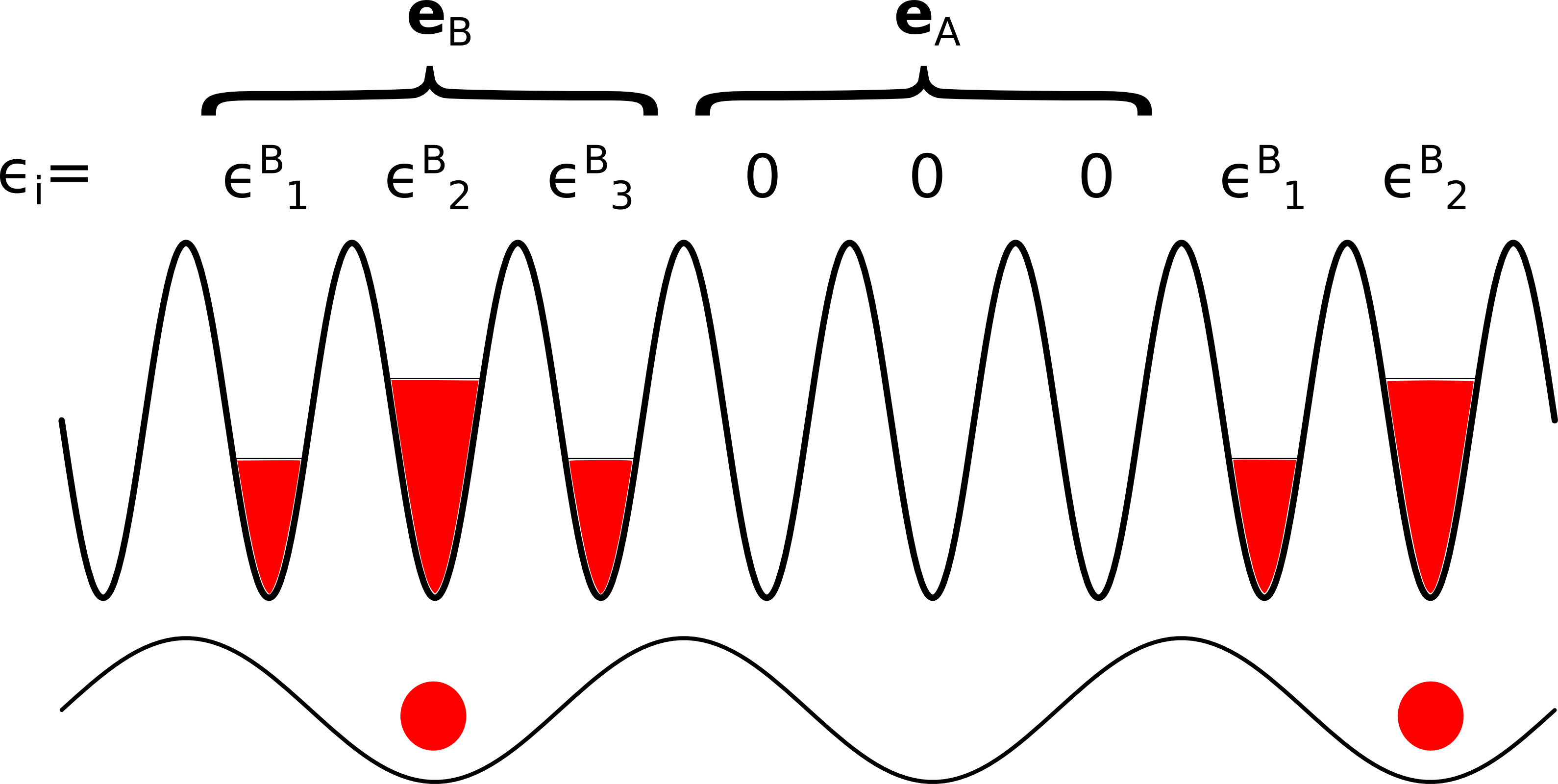}
  \caption{Setup for gnmer (with length $l=b^f/b=3$): two overlapping optical lattices (shifted on picture). The first lattice affects the \emph{mobile} particles, latter the \emph{frozen} ones. If the interactions between the species falls with the distance fast enough, we can decompose such a system into two types of blocks of length $l$. When the \emph{frozen} particle is absent $l$ corresponding sites of main lattice have zero on-site energy, when it is present (depicted as red disc), it effectively changes energies in $l$ sites of the main lattice in a fixed way. \label{pic:GNMER2}}
 \end{center}
\end{figure}
\paragraph{Generalized N-mers}
(gnmers) can be created in setups with $b^f=l b$ where $l\in\mathbb{N}$. In such a case one site for the \emph{frozen} atoms range over $l$ sites for the \emph{mobile} ones and we can distinguish two types of structures (as on fig. \ref{pic:GNMER2}):
\begin{itemize}                                                                                                                                                                                                                                                       \item empty rows ($T_A(E)=T^l_0(E)$),
\item structures with inhomogeneous set of energies $\mathbf{e}_B$ and/or tunnelings $\mathbf{t}_B$ ($T_B(E)$).
\end{itemize}
As $T_A(E)$ describes an empty row -- special case of $N$-mer -- it will provide $l-1$ resonances \eqref{eq:resnmer} (albeit some of them may be suppresed due to falling out of band for structure $B$).

The structure of the block $B$ is crucial for the existence of additional resonances. We will focus on examples possible to create in the presented setup. An interaction with a \emph{frozen} atom will be the strongest at the central site of the structure and it will fall towards the edges of block. We assume that the interactions become negligible for distances bigger than the width of one block.
In this scope $N$-mers presented in the preceding paragraph could be treated as one limiting case of gnmers: with interactions constant for the whole range of one block (square wall), which is rather unphysical setting.

\begin{figure}
 \begin{center}
  \includegraphics[width=0.5\textwidth]{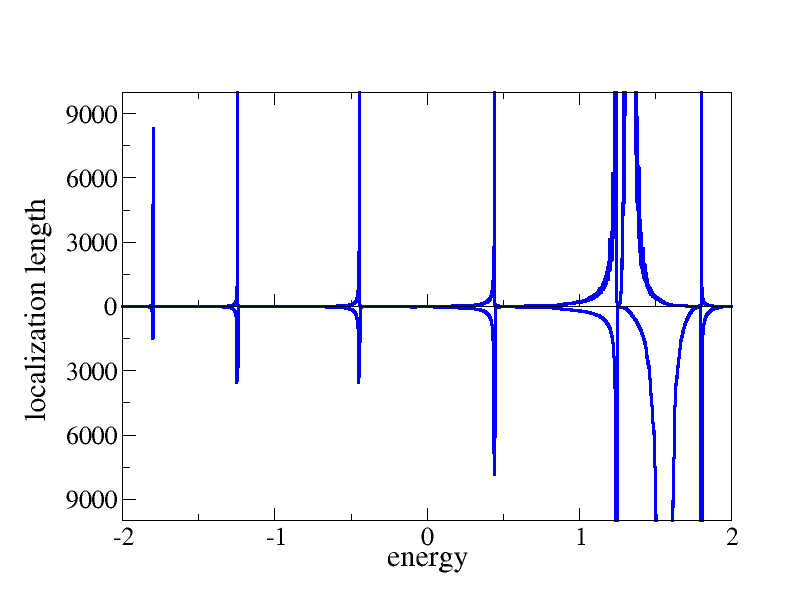}
  \caption{The localization length in function of the energy calculated numerically by the transfer matrix method for gnmer with $T_A=T^l_0(E)$ and $T_B=T_{\mathrm{sr}}(V,t',l,E)$ for $l=7$, $V=1$. Upper pane shows results for $t'=0.5$, lower for $t'=0.6$. Resonances are very narrow and asymmetric as most of them lies at the edges of gaps in spectrum of $T_\mathrm{sr}$, one resonance changes its position due to change of $t'$. \label{pic:sr_all}}
 \end{center}
\end{figure}
The opposite case (yet physical) are short ranged interactions $(sr)$, when interaction between the \emph{frozen} and the \emph{mobile} particles is non-zero only in one central site of the block. Thus $\mathbf{e}_\mathrm{sr}$ is defined by $\epsilon_{0}^\mathrm{sr}=V$ and $\epsilon_i^\mathrm{sr}=0$ for $i\neq0$ (for sake of clearness, from now on we will use only gnmers with odd $l$ and adopt intuitive numbering with $i=0$ in central site of the block and $i\in[-(l-1)/2,(l-1)/2]$.). For this case
\begin{align}
T_\mathrm{sr}(V,l,E)=T_0^{l}(E)+T_0^{(l-1)/2}(E)\cdot\left(\begin{array}{cc}V&0\\ 0&0\end{array}\right)\cdot T_0^{(l-1)/2}(E). 
\end{align}
It is straightforward to show using recurrence properties of Chebyshev polynomials that \eqref{eq:rescond} for $T_\mathrm{sr}$ is satisfied only for $V=0$. Thus in this simple case we do not have any additional resonant energies.
In this setup an additional extended mode could be created by changing the tunneling amplitudes by using the fast periodic modulation of interaction strength as mentioned in the model description. In this case it will result in changing $t_i$ only around central site $i=0$ (namely $t_0^\mathrm{sr}=t_{-1}^\mathrm{sr}=t',$ where $t'=\mathcal{J}_0(\delta V/\omega)$ and $\delta V$ is amplitude of interaction strength modulation and $\omega$ modulation frequency), the rest of tunnelings are unaffected. We denote transfer matrix for this kind of block by $T_\mathrm{sr'}(V,t',l,E)$. For this model the extended state appears (same as in DRDM model \cite{vignolo2010,kosior15,kosior15b}) with $E_R=V/(1-t'^2)$.
In the fig.\ref{pic:sr_all} we present the localization length calculated numerically using the standard transfer matrix method \cite{delande2011} (the localization length is given in units of a lattice constant i.e. a distance between $j$-th and $i$-th site is $|i-j|$). The results are shown for $T_B=T_{\mathrm{sr}}(V,t',l,E)$ with $V=1$, $l=7$ and for two different $t'\in\{0.5,0.6\}$ (upper/lower pane). All predicted extended  modes are present, one of them (created by the presence of $t'\neq1$) can be moved relatively to others. 
The chosen parameters could be attained in an experiment. The necessity of using two optical lattices with lattice constants differing by a large factor (in our case $7$) seems to be most demanding experimental requirement, but it could be met (even solely in infrared), with the use of CO2 laser (with wavelength $10.6\mathrm{\mu m}$) \cite{Friebel98} and lasers with frequencies from telecommunication band ($1260$-$1675\mathrm{nm}$) or similar \cite{Haller2010}.

\begin{figure}
 \begin{center}
  \includegraphics[width=0.5\textwidth]{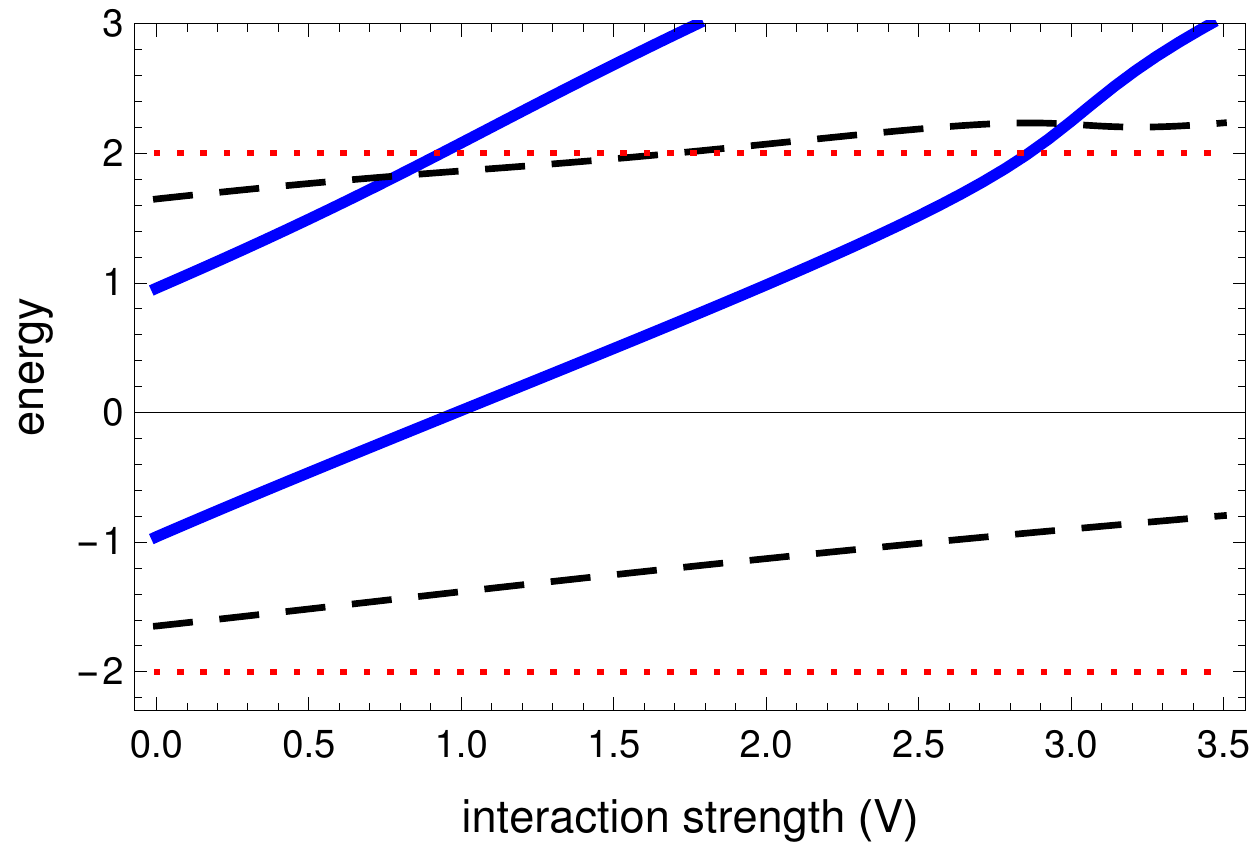}
  \caption{Solutions of equation \eqref{eq:rescond} for $T_B=T_{\mathrm{dim}}(V,d,l,E)$ for $l=7$ and $d=0.9$ in function of energy and interaction strength $(V)$. Blue solid lines represents real solutions -- for those parameters extended states exists, black dashed lines are complex solutions, in vicinity of them localization length could be increased (especially if imaginary component is small) however is finite, red dotted lines are band edges for $T^A=T^l_0$.\label{pic:resdim}}
 \end{center}
\end{figure}
\begin{figure}
 \begin{center}
  \includegraphics[width=0.5\textwidth]{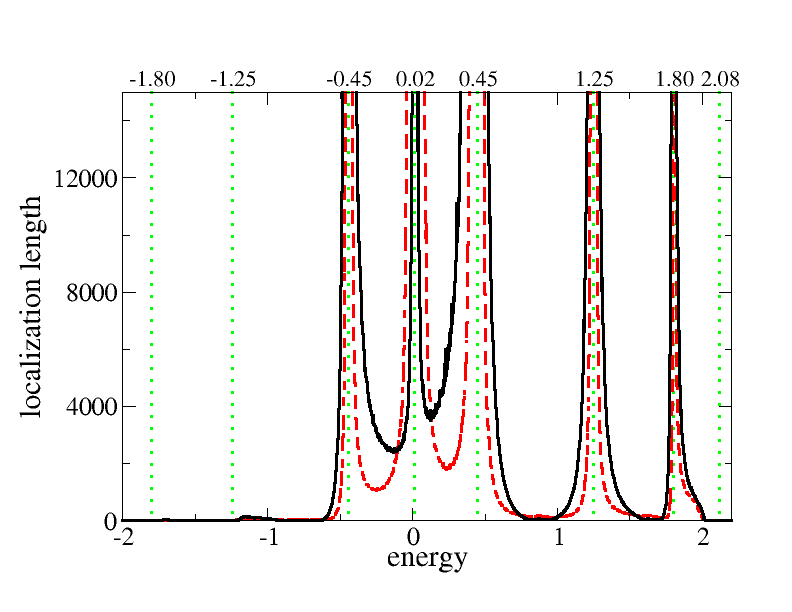}
  \caption{The localization length in function of the energy calculated numerically by the transfer matrix method. Black line is the result of disorder in form of gnmer with $T_A=T^7_0(E)$ and $T_B=T_{\mathrm{dim}}(V,d,l,E)$ for $l=7$, $d=0.9$ and $V=1$. Red dashed line is the result of the same system but without interaction range cutoff assumed in the case of gnmer. Green dotted vertical lines represent predicted resonances, some of them are out of band.\label{pic:dim_all}}
 \end{center}
\end{figure}
For cases of exponentially or polynomially falling interactions, resonant energies can be found by numerical solution of \eqref{eq:rescond} for given $T_B$. As for most of the investigated models results were qualitatively similar, we present only one case of interactions falling with the third power of distance, namely $\epsilon_0^B = V$, $\epsilon_i^B = \frac{V_d}{|i|^3}$ for $i\neq0$.
Resonant energies calculated for the blocks with  $T_B=T_\mathrm{dim}(V,V_d,l,E)$ ($l=7$ and $V_d/V=0.9$) are plotted on fig. \ref{pic:resdim}, solid lines represent resonant energies (real solutions), while dashed denotes complex solutions.
Results of numerical calculations of localization length for this model with $V=1$ are shown in fig \ref{pic:dim_all}.
In this system most of resonances appears on predicted positions. Three of the resonances are absent as they are lying out of bands. The resonance for $E\approx 0.02$ could be moved relatively to rest by changing $V$ or $V_d$.

Finally we will comment on the assumption that blocks are separable. The situation is clear for short ranged interactions, moreover it automatically satisfies another assumption we made -- that tunnelings at block edges are always the same. It is so as tunneling amplitudes depend on the difference of on site energies, thus if on site energies have negligible values, the change of tunneling can also be neglected.

For long range interactions we could check if loosing assumption that blocks are separable qualitatively change numerical results. For the case presented above, we calculated propagation of particle in potential with the same cubic decay and the same parameters as for $\mathbf{e}_\mathrm{dim}$ but without the cutoff on block edges (practically in calculations cutoff was set for interaction strength smaller than $2\cdot10^{-5}$). Results are plotted on fig \ref{pic:dim_all} with red dashed line. Although some features differ and we could not prove that resonances are infinite, qualitatively the  plots are the same. It shows that at least in some cases resonances do not vanish immediately if interactions are long ranged and assumption of block separability is not well preserved.
\section{Conclusions \label{sec:concl}}
We described the one dimensional tight-binding model with the correlated disorder of specific type, created by using two species of ultracold atoms in two optical lattices with perpendicular polarization. In such a setup, if lattice constants of two lattices are commensurable (but not equal), the correlations of a type of generalized $N$-mer appear --the system is composed of randomly ordered blocks of two types. For such a systems, there can exist extended states and we devised a simple formula allowing the determination of their energies.
We showed that for experimentally attainable parameters there exists systems with multiple extended modes and confronted analytical predictions with numerical calculations of localization length for such systems. Furthermore, we checked that even for long ranged interactions, cutting interaction range at the edge of one block gives qualitatively good results. Presented systems could be used as tunable band-pass filters for particle energies, trapping all atoms but those with very specific energies, thus allowing creation of a multi-mode ''gun'' for matter waves. The general scheme presented in section \ref{sec:method} could also be used to engineer systems with narrow bands of conductance in other media such as optical waveguides \cite{Somekh73} or semiconductor nanostructures \cite{Beenakker1991}.

\section{Acknowlegnments}
We are grateful to A. Kosior, M. P\l odzie\'n, K. Sacha and K. Zakrzewski for discussions and critical reading of the manuscript, also to anonymous referee for comments which helped a lot in improving the manuscript.  We acknowledge support of the Polish National Science Center via Project Preludium No. 2015/19/N/ST2/01677 and of EU via project QUIC (H2020-FETPROACT-2014 No.641122).


\end{document}